\documentstyle[11pt,epsf]{article}
\newcommand{\beq}{\begin{equation}}
\newcommand{\eeq}{\end{equation}}
\newcommand{\beqa}{\begin{eqnarray}}
\newcommand{\eeqa}{\end{eqnarray}}
\newcommand{\beqan}{\begin{eqnarray*}}
\newcommand{\eeqan}{\end{eqnarray*}}
\newcommand{\no}{\nonumber}

\newcommand{\ul}{\underline}
\newcommand{\ol}{\overline}
\newcommand{\ra}{\rightarrow}

\newcommand{\ben}{\begin{enumerate}}
\newcommand{\een}{\end{enumerate}}
\newcommand{\bfl}{\begin{flushleft}}
\newcommand{\efl}{\end{flushleft}}
\newcommand{\ba}{\begin{array}}
\newcommand{\ea}{\end{array}}
\newcommand{\btab}{\begin{tabular}}
\newcommand{\etab}{\end{tabular}}
\newcommand{\bit}{\begin{itemize}}
\newcommand{\eit}{\end{itemize}}

\newcommand{\vs}{\vspace}
\newcommand{\hs}{\hspace}

\newcommand{\G}{\Gamma}
\newcommand{\prepr}[1] {\begin{flushright} {\bf #1} \end{flushright} \vskip
1.cm}
\newcommand{\titul}[1] {\begin{center}{\Large {\bf #1 } } \end{center}
\vskip 0.8cm}

\newcommand{\autor}[1] {\begin{center} \large {\bf \lineskip .3cm #1  }
                        \end{center} }

\newcommand{\lugar}[1] {\begin{center}  {\normalsize \bf \it #1   }
\end{center}}
\newcommand{\abstr}[1] {{\begin{center} \vskip .5cm {\bf \large Abstract
                        \vspace{0pt}} \end{center}}\begin{quote} \small #1
                        \end{quote}}
\newcounter{muni}

\input epsf
\begin{document}
\vspace{.1cm}
\hbadness=10000
\pagenumbering{arabic}
\begin{titlepage}
{\bf \em 6 March, 1995} \hs{80mm}
\prepr{Preprint PAR/LPTHE/95-09 \hs{4mm} \\ hep-ph/9503326 }
\titul{\large A MODEL FOR THE DECAY OF THE $D_s^{+}$ MESON INTO THREE PIONS}
\autor{ M. Gourdin\footnote{\rm Postal address: LPTHE, Tour 16, $1^{er}$ Etage,
4 Place Jussieu, F-75252 Paris CEDEX 05, France.},
Y. Y. Keum\footnote{\rm Postal address: LPTHE, Tour 24, $5^{\grave{e}me}$
Etage,
 2 Place Jussieu, F-75251 Paris CEDEX 05, France. \\
{\small  E-mail : gourdin@lpthe.jussieu.fr, keum@lpthe.jussieu.fr,
 and pham@lpthe.jussieu.fr.}
}
and X. Y. Pham$^*$ }

\lugar{Universit\'e Pierre {\it \&} Marie Curie, Paris VI \\
Universit\'e Denis Diderot, Paris VII \\
Physique Th\'eorique et Hautes Energies, \\
Unit\'e associ\'ee au CNRS : D 0280
}

\vs{-14cm}
\thispagestyle{empty}
\vs{133mm}
\noindent
\abstr{
We propose a phenomenological two component model
describing the decay amplitude of the process
$D_s^{+} \ra 3 \pi$, whose rate has been found surprisingly large.
The first component is a constant background  $F_{NR}$,
and the second one is a Breit-Wigner type amplitude
associated to a quasi two body $f_0(980)$ $\pi^{+}$ state.
We show that it is possible to reproduce the observed rate
for $D_s^{+} \ra \pi^{+}\pi^{+}\pi^{-}$
as well as the two other measured branching ratios
for the non resonant part and the  resonant $f_0\pi^{+}$ one,
with a common parameter $F_{NR}$.

Predictions are given for
the $D_s^{+} \ra  \pi^{0}\pi^{0}\pi^{+}$ rates,
as well as for the various $\pi^{+}$ and $\pi^{-}$,
or $\pi^{0}$ and $\pi^{+}$ energy distributions
for the two possible final states.

}
%

\end{titlepage}


\newpage
\hspace{1cm} \large{} {\bf I \hs{3mm} Introduction}    \vspace{0.5cm}
\normalsize

We are interested, in this paper,
in the decay of the $D_{s}^{+}$ meson
into three pions.
We observe that both flavoured constituents  c and $ \ol{s}$
of the $D_{s}^{+}$ meson are absent in the final state.
As a consequence, the decay mode $D_{s}^{+}$ $\ra$ 3 $\pi$ can
be described  neither by a spectator, nor by a colour suppressed,
nor by a penguin diagram, but only by a $W$ annihilation
where the virtual time-like $W^{+}$ decays into three pions.

It is generally expected
that $W$ annihilation amplitudes are small compared to
spectator and colour suppressed ones.
Surprisingly experimental data do not seem
to support this belief. As an illustration,
let us consider the three following decay modes of the $D_{s}^{+}$ meson
involving the same Cabibbo, Kobayashi, Maskawa (CKM) factor
$V_{cs}^{*}  V_{ud}$ and compare their branching ratios \cite{R1} :

\hs{20mm}  i) \hs{3mm} Spectator ;
$$
Br(D_{s}^{+} \ra \phi + \pi^{+} ) = (3.5 \pm 0.4) \hs{2mm} \% \hs{3mm},
$$

\hs{20mm} ii) \hs{2mm} Colour suppressed ;
$$
Br(D_{s}^{+} \ra \ol{K}^{0} + K^{+} ) = (3.5\pm 0.7) \hs{2mm} \% \hs{3mm},
$$

\hs{20mm} iii) \hs{1mm} $W$ annihilation ;
 $$
Br(D_{s}^{+} \ra \pi^{+} + \pi^{+} + \pi^{-} )
= (1.35 \pm 0.31) \hs{2mm} \% \hs{3mm}.
$$
It is therefore interesting to understand the origin for
such a large value of the $D_{s}^{+} \ra  3 \pi$ branching ratio.

The transition of the virtual $W$ into three pions
occurs also in the $\tau$ decay mode
$\tau^{+} \ra \ol{\nu}_{\tau} + 3 \pi$.
However, in this case, the full axial vector weak
current contributes,
and the amplitude depends on three independent structure functions.
For the decay $D_{s}^{+} \ra  3 \pi$,
only one structure function associated to the divergence
of the axial vector current is involved,
because here the $W$ is in the $J^{P} = O^{-}$ state,
like the $D_{s}$ meson.
Therefore the study of $D_{s} \ra 3 \pi $ decay
is the cleanest way to isolate the structure function
$F_{4}$ of references \cite{R2, R3}.

In Section II, we develop the formalism for $D_{s}^{+}  \ra 3 \pi$ decay.
We study the kinematics and the three pion phase space.
Experimental data as quoted in Reference \cite{R1} are listed.

By inspection of these experimental data,
we observe that the quasi two body state $\rho^{o} \pi^{+}$ assumed in
\cite{R2, R3} to be responsible of the structure function $F_{4}$
 is negligible compared to the quasi two body state $f_{0}(980) \pi^{+}$.
Moreover, since the non resonant rate is large,
 we propose a model in Section III where the decay
amplitude for $D_{s}^{+}  \ra  3 \pi$ has two components :
the first one is associated to
a background non resonant three pions,
and the second one corresponds to a quasi two body
$f_{0}$ $\pi^{+}$ final state.

We compare our model with experiment in Section IV for the final state
$\pi^{+} \pi^{+} \pi^{-}$ where data on rates are available
and we make predictions
for the $\pi^{+}$ and $\pi^{-}$ energy distributions.

For the final state $\pi^{0} \pi^{0} \pi^{+}$,
using isospin analysis, we make predictions in Section V
for the rates and the $\pi^{0}$ and $\pi^{+}$
energy distributions.

Our model is consistent with the rate measurements.
However the serious test would be
the observation of the various $\pi$ meson energy distributions and ultimately,
if statistics is copious enough, of the Dalitz plots.

\vs{7mm}
\large{}
\hs{10mm} {\bf II \hs{3mm} Generalities and Kinematics}
\vspace{0.5cm}
\normalsize

We study the decay of a $D_s^{+}$ meson of energy momentum $P$
into three pions of energy momenta $p_1, p_2, p_3$ with
the relation $P = p_1 + p_2 + p_3$.
We introduce the Mandelstam variables $s_1, s_2, s_3$ and the $\pi$
meson energies $E_1, E_2, E_3$ in the $D_s^{+}$ rest frame.
Neglecting the mass difference between charged and neutral pions,
we get
\beqa
s_1 &=& (p_2 + p_3)^2 = (P - p_1)^2
= m_{D_s}^2 + m_{\pi}^2 - 2 m_{D_s} E_1 \hs{3mm}, \no \\
s_2 &=& (p_3 + p_1)^2 = (P - p_2)^2
= m_{D_s}^2 + m_{\pi}^2 - 2 m_{D_s} E_2 \hs{3mm}, \label{eq:2-1} \\
s_3 &=& (p_1 + p_2)^2 = (P - p_3)^2
= m_{D_s}^2 + m_{\pi}^2 - 2 m_{D_s} E_3 \hs{3mm}. \no
\eeqa

Energy momentum conservation implies the relations
\beq
s_1 + s_2 + s_3 = m_{D_s}^2 + 3 m_{\pi}^2 \hs{15mm},
\hs{15mm} E_1 + E_2 + E_3 = m_{D_s}  \hs{3mm}. \label{eq:2-2}
\eeq

The double differential distribution is given by
\beq
d\G = \frac{1}{64 \hs{2mm} \pi^3} \hs{2mm}
\frac{1}{m_{D_s}} \
\left| < 3 \pi | T | D_s^{+} > \right|^2 \hs{2mm}
dE_1 \hs{2mm} dE_2 \hs{3mm}, \label{eq:2-3}
\eeq
and the transition matrix element $<3 \pi| T | D_s^{+} >$
involving only spinless particles is dimensionless.

In the $(E_1, E_2)$ plane, the phase space is defined by the constraints
\beqa
m_{\pi} \hs{2mm} \leq \hs{2mm} & E_1 & \hs{2mm} \leq
\hs{2mm} \frac{m_{D_s}^2 - 3 m_{\pi}^2}
{2 \hs{2mm} m_{D_s}} \hs{3mm}, \no \\
E_{-}(E_1) \hs{2mm} \leq \hs{2mm} & E_2 & \hs{2mm} \leq  \hs{2mm}
E_{+}(E_1) \hs{10mm}, \label{eq:2-4}
\eeqa
with
\beq
E_{\pm}(E) = \frac{1}{2} (m_{D_s} - E) \pm \frac{1}{2}
\hs{2mm} \left\{ \frac{(E^2 - m_{\pi}^2)(m_{D_s}^2 - 3 m_{\pi}^2 - 2 m_{D_s}
E)}
{m_{D_s}^2 + m_{\pi}^2 - 2 m_{D_s} E} \right\}^{1/2} \hs{3mm}.
\label{eq:2-5}
\eeq

Of course, the mass difference between charged and neutral pions
being neglected, we have the same phase space in the two other planes :
 $(E_2, E_3)$ and $(E_1, E_3)$.

At the quark level, the decay $D_s^{+} \ra 3 \pi$ is described by a
\hs{0.5mm} $W$
annihilation diagram,
and the transition matrix element is given by
\beq
< 3 \pi| T | D_s^{+} > \hs{2mm} = \hs{2mm}
\frac{G_F}{\sqrt{2}} \hs{2mm} V_{cs}^{*}
\hs{2mm} V_{ud} \hs{2mm} a_1 \hs{2mm} f_{D_s} \hs{2mm}
P^{\mu} \hs{2mm} < 3 \pi | A_{\mu} | 0 > \hs{3mm}. \label{eq:2-6}
\eeq
where $a_1$ is the phenomenological parameter introduced by
Bauer, Stech, Wirbel \cite{R4}, and
$f_{D_s}$ is the leptonic decay constant of the $D_s$ meson.
The matrix element of the divergence of the weak axial
current between the vacuum and three pion final state
is an unknown form factor $F(E_1, E_2)$,
function of two independent variables,
taken as the pion energies.
\beq
P^{\mu} \hs{2mm} < 3 \pi | A_{\mu} | 0 > \hs{2mm} = \hs{2mm}
m_{D_s} \hs{2mm} F(E_1, E_2) \hs{3mm}. \label{eq:2-7}
\eeq
Since the transition matrix element is dimensionless,
the form factor $F(E_1, E_2)$ \footnote{Our $F(E_1, E_2)$ is related to
the $F_4$ of the Ref.\cite{R2} by
$F(E_1, E_2) = m_{D_s} \hs{1mm}
F_4(s_1, s_2, Q^2 = m_{D_s}^2)$ } is also dimensionless.

In the $(E_1, E_2)$ plane, the Dalitz plot is given by :
\beq
d\G(D_s^{+} \ra 3 \pi) = \frac{m_{D_s}}{64 \hs{1mm} \pi^3} \hs{2mm}
\left( \frac{G_F}{\sqrt{2}} \right)^2 \hs{2mm} |V_{cs}|^2 \hs{2mm}
|V_{ud}|^2 \hs{2mm} a_1^2 \hs{2mm} f_{D_s}^2 \hs{2mm}
|F(E_1, E_2)|^2 \hs{2mm} dE_1 \hs{2mm} dE_2 \hs{3mm}. \label{eq:2-8}
\eeq
The decay rate is obtained by integration of the distribution
(\ref{eq:2-8}) over the energies $E_1$ and $E_2$ inside
the phase space described in Eq.(\ref{eq:2-4}).
The branching ratio can  be written in the form :
\beq
Br(D_s^{+} \ra 3 \pi) = \frac{\tau_{D_s^+} \hs{1mm} m_{D_s}}{\hbar} \hs{2mm}
\frac{1}{64 \hs{1mm} \pi^3}
\left[ \frac{G_F \hs{1mm} m_{D_s}^2}{\sqrt{2} } \right]^2
\hs{2mm} |V_{cs}|^2 \hs{2mm}
|V_{ud}|^2 \hs{2mm} a_1^2 \hs{2mm}
\left( \frac{f_{D_s}}{m_{D_s}} \right)^2 \hs{2mm} I \hs{3mm}, \label{eq:2-9}
\eeq
where the dimensionless integral $I$ is defined by :
\beq
I = \frac{1}{m_{D_s}^2} \int \int \hs{2mm}
|F(E_1, E_2)|^2 \hs{2mm} dE_1 \hs{2mm} dE_2 \hs{3mm}.  \label{eq:2-10}
\eeq

Using $\tau_{D_s^{+}} = 4.67 \cdot 10^{-13} s$ \cite{R1}
and $f_{D_s} = 280 MeV$
\footnote{The leptonic decay  $D_s^{+} \ra \mu^{+} + \nu_{\mu}$
has been measured by two groups and the extracted values of $f_{D_s}$
are $f_{D_s} = (232 \pm 69) \hs{2mm} MeV$ and
$f_{D_s} = (344 \pm 76) \hs{2mm} MeV$.
Our choice of $f_{D_s} = 280 \hs{2mm} MeV$ is
consistent with  data and with theoretical expectations. },
we get
\beq
Br(D_s^{+} \ra 3 \pi) = 1.31 \hs{1mm} \cdot \hs{1mm} 10^{-2}
\hs{2mm} a_1^2 \hs{2mm} I \hs{3mm}. \label{eq:2-11}
\eeq

{}From the analysis of colour favoured $D$ meson decay,
a reasonable value for $a_1$ is $a_1 = 1.26$ \cite{R5} and we obtain
\beq
Br(D_s^{+} \ra 3 \pi) = 2.08
\hs{1mm} \cdot \hs{1mm} 10^{-2} \hs{2mm} I \hs{3mm}. \label{eq:2-12}
\eeq
We make the following choice of $\pi$-meson energy variables ;

\hs{20mm} i) \hs{2mm}
final state $\pi^{+} \pi^{+} \pi^{-}$ \hs{20mm}
$E_1(\pi^{+}), E_2(\pi^{+}), E_3(\pi^{-})$ \hs{3mm},

\hs{20mm} ii) \hs{2mm}
final state $\pi^0 \pi^0 \pi^{+}$ \hs{20mm}
$E_1(\pi^0), E_2(\pi^0), E_3(\pi^{+})$ \hs{3mm}.

Due to Bose-Einstein symmetry, the function $F(E_1, E_2)$ is
symmetrical in the exchange between $E_1$ and $E_2$.
Such a property extends to the Dalitz plot given by
the double differential distribution represented by
$|F(E_1, E_2)|^2$.
It is probably premature to discuss the details of the Dalitz plot
and it might be interesting to define one meson energy distributions :
\beqa
G(E_1) &=& \frac{1}{m_{D_s}} \int^{E_{+}(E_1)}_{E_{-}(E_1)}
\hs{2mm} |F(E_1, E_2)|^2 \hs{2mm} dE_2 \hs{3mm}, \label{eq:2-13} \\
H(E_3) &=& \frac{1}{m_{D_s}} \int^{E_{+}(E_3)}_{E_{-}(E_3)}
\hs{2mm} |F(m_{D_s} - E_2 - E_3, E_2)|^2 \hs{2mm} dE_2 \hs{3mm},
\label{eq:2-14}
\eeqa
where
$G(E_1)$ corresponds to the $\pi^{+} (\pi^0)$ for the final state
$\pi^{+}\pi^{+}\pi^{-} (\pi^0\pi^0\pi^{+})$,
and $H(E_3)$ corresponds to the $\pi^{-} (\pi^{+})$  for the final state
$\pi^{+} \pi^{+} \pi^{-} (\pi^0 \pi^0 \pi^{+})$.

The presently available data concern the rates for the decay mode
$D_s^{+} \ra \pi^{+} \pi^{+} \pi^{-}$.
They are given in Table 1, the last column indicates
the value of the quantity $I$ determined from experiment \cite{R1}
by using Eq.(\ref{eq:2-12}).

\vs{3mm}
\begin{table}[thb]
\begin{center}
\begin{tabular}{|c||c||c||}
\hline
Mode
& Experimental  Branching Ratios
& Experimental  values of $I$ \\
\hline
\hline
$ \pi^{+}\pi^{+}\pi^{-}$ &
$(1.35 \pm 0.31) \cdot 10^{-2} $ &
$0.649 \pm 0.149 $ \\
\hline
$ (\pi^{+}\pi^{+}\pi^{-})_{NR}$ &
$(1.01 \pm 0.35) \cdot 10^{-2} $ &
$0.486 \pm 0.168 $ \\
\hline
$ f_{0} \pi^{+}$ &
$(1 \pm 0.4) \cdot 10^{-2} $ &
$0.481 \pm 0.192 $ \\
\hline
$ \rho^0 \pi^{+}$ &
$ \leq 0.28 \cdot 10^{-2}  $ &
$ \leq 0.135 $ \\
\hline
\hline
\end{tabular}

\vs{3mm}
Table 1 \\
\end{center}
\end{table}

\vs{3mm}
\hs{3mm} \large{ } {\bf III. \hs{3mm} Phenomenological Model for $D_s^{+} \ra 3
\pi$ }
\normalsize
\vs{3mm}

In order to describe the decay mode $D_s^{+} \ra 3 \pi$,
we use a simple phemonenological model where the function $F(E_1, E_2)$
is the sum of two contributions ;
\beq
F(E_1, E_2) = F_{NR} + F_{RES}(E_1, E_2) \hs{3mm}. \label{eq:3-1}
\eeq
Obviously, this decomposition is guided by experimental data \cite{R1}.
The first component $F_{NR}$ is assumed to be a real constant describing
the non resonant part of the amplitude.
The second component $F_{RES}(E_1, E_2)$ is associated to the quasi
two body final state $D_s^{+} \ra f_{0} + \pi^{+}$ followed
by the subsequent decays
$f_{0} \ra \pi^{+}\pi^{-}$ or $\pi^0 \pi^0$.

\large
1). \normalsize
At first, consider the charged case $f_{0} \ra \pi^{+}\pi^{-}$.
The $\pi^{-}$ meson has an energy $E_3$ and the two $\pi^{+}$ mesons
energies $E_1$ and $E_2$.
We then have two possible resonant combinations and the general form
of $F_{RES}(E_1, E_2)$ is :
\beq
F_{RES}(E_1, E_2) \hs{2mm} = \hs{2mm}
D_c \hs{2mm} \left\{BW(E_1) + BW(E_2) \right\} \hs{2mm},  \label{eq:3-2}
\eeq
where the dimensionless Breit-Wigner function $BW(E_j)$ is defined by :
\beq
BW(E_j) \hs{2mm} = \hs{2mm}
\frac{ m^2_{f_{0}}}{m^2_{f_{0}} - s_{j} - i \sqrt{s_j}
\hs{2mm} \G_{f_{0}}} \label{eq:3-3}
\eeq
with $s_j$ and $E_j$ related by
$s_j = m_{D_s}^2 + m_{\pi}^2 - 2 m_{D_s}E_j$ ( $j $ = 1, 2).

The complex constant $D_c$ corresponds to the transition
$W^{+} \ra f_{0} \pi^{+}$,
for which we assume,
by the Partial Conservation of the Axial Current ($PCAC$)
in Eqs.(\ref{eq:2-6}) and (\ref{eq:2-7}),
the existence of an intermediate state having the quantum number of
a $\pi^{+}$ meson \cite{R2, R3, R6}
and which might be the $\pi$ meson  itself
or its recurrence $\pi(1300)$.
It turns out that the $\pi$ meson intermediate state
will give a contribution many order of magnitude smaller than
that of the $\pi(1300)$ and only the later one is retained
given the following expression for $D_c$
\footnote{ We have used for width the simple energy dependence
$\G(s) = \frac{\sqrt{s}}{m} \G(m)$.
In the case of the $\pi(1300)$ width, more sophisticated dependence
has been proposed \cite{R2}.
Because of the not too large difference between the $\pi(1300)$ and $D_s^{+}$
masses, we expect the sensitivity of Eq.(\ref{eq:3-4})
to different forms of $\G_{\pi^{'}}(s)$ to be relatively modest. }
:
\beq
D_c \hs{2mm} = \hs{2mm}
\frac{m_{D_s}}{m_{f_{0}}} \hs{1mm}
\frac{f_{\pi^{'}} \hs{2mm} m_{\pi^{'}}}
{m_{\pi^{'}}^2 - m_{D_s}^2 - i \hs{1mm} m_{D_s} \G_{\pi^{'}}}
\hs{2mm} g_{\pi^{'}f_{0} \pi}
\hs{2mm} g_{f_{0} \pi^{+} \pi^{-}}  \hs{2mm}. \label{eq:3-4}
\eeq

In Eq.(\ref{eq:3-4}), $\pi^{'} \equiv \pi(1300)$,
and the dimensionless coupling constants $g_{M_0 M_1 M_2}$
describe the decay of a spinless meson $M_0$ into two spinless mesons
$M_1$ and $M_2$ :
\beq
< M_1 \hs{2mm} M_2 \hs{2mm}| \hs{2mm} T \hs{2mm} | \hs{2mm} M_0 >
\hs{2mm} = \hs{2mm} m_0 \hs{2mm} g_{M_0 M_1 M_2} \hs{3mm}. \label{eq:3-5}
\eeq
The term $m_{D_s}$ in the numerator of Eq.(\ref{eq:3-4}) comes from
the definition Eq.(\ref{eq:2-7}),
the term $m_{f_{0}}$ in the denominator of Eq.(\ref{eq:3-4})
when combined with Eq.(\ref{eq:3-3}) yields the definition
Eq.(\ref{eq:3-5}).
Also $m_{\pi^{'}}$ comes from Eq.(\ref{eq:3-5}) and $f_{\pi^{'}}$ comes
from the coupling between $\pi^{'}$ and $W$.
The numerical value of $g_{M_0M_1M_2}$ is obtained
from the experimental decay rate $ \G(M_0 \ra M_1 + M_2)$
by the relation
\beq
g^2_{M_0 M_1 M_2} \hs{2mm} = \hs{2mm}
8 \hs{1mm} \pi \frac{\G(M_0 \ra M_1 + M_2)}{K} \hs{3mm}, \label{eq:3-6}
\eeq
where $K$ is the final momentum in the $M_0$ rest frame.

Now we consider first the case $f_{0} \ra 2 \pi$.
Using the experimental parameters \cite{R1} :
\beqa
& & m_{f_{0}} = 980 \hs{2mm} MeV \hs{10mm}, \hs{10mm}
\G_{f_{0}} = (49 \pm 9) \hs{2mm} MeV \hs{3mm}, \no \\
& & Br(f_{0} \ra 2 \pi) \hs{2mm} =
\hs{2mm} 0.781 \hs{1mm} \pm \hs{1mm} 0.024 \hs{3mm}, \label{eq:3-7}
\eeqa
we obtain
\beq
g_{f_{0} \pi^{+}\pi^{-}} = 1.144 \pm 0.109 \hs{5mm}, \hs{5mm}
g_{f_{0} \pi^{0}\pi^{0}} = 0.809 \pm 0.077  \hs{3mm}. \label{eq:3-8}
\eeq
The isoscalar property of the $f_{0}$ meson has been taken into
account for relating both coupling constants.

Secondly, for the decay  $\pi(1300) \ra f_{0} + \pi$,
experimental data are poor \cite{R1} and we shall use :
\beqa
& & m_{\pi^{'}} = 1300 \hs{2mm} MeV \hs{10mm}, \hs{10mm}
\G_{\pi^{'}} = (400 \pm 200) \hs{2mm} MeV \hs{3mm}, \no \\
& & Br(\pi^{'} \ra f_{0} \pi) \hs{2mm} =
\hs{2mm} 0.68 \hs{3mm}. \label{eq:3-9}
\eeqa
The result is
\beq
g_{\pi^{'} f_{0} \pi} \hs{2mm} = \hs{2mm}
5.2 \hs{1mm}^{+ \hs{1mm} 1.2}_{- \hs{1mm} 1.5} \hs{3mm}. \label{eq:3-10}
\eeq
The last parameter entering in Eq.(\ref{eq:3-4}) is the leptonic
constant $f_{\pi^{'}}$.
The estimate of reference \cite{R6} is
\footnote{ Different values of $f_{\pi^{'}}$ have been proposed
in the literature.
See for instance references \cite{R2} and \cite{R3}.}
\beq
f_{\pi^{'}} \hs{2mm} = \hs{2mm} 40 \hs{2mm} MeV \label{eq:3-11}
\eeq

Now we are in a position to compute the numerical value
of the dimensionless coupling constant $D_c$
written in the form $D_c = |D_c|$ $exp(i \hs{1mm} \Phi_D)$.
Retaining only the large error due to the poor knowledge of
the total $\pi(1300)$ width, we obtain for different
$\G_{\pi^{'}}$ values :
\beq
\G_{\pi^{'}} = \left| \begin{array}{c} 600 \\ 400 \\ 200  \end{array}
\right| \hs{2mm} MeV
\hs{4mm}, \hs{4mm}
|D_c| \hs{2mm} = \hs{2mm}
\left| \begin{array}{c} 0.3068 \\ 0.2679 \\ 0.1981 \end{array}
\right|
\hs{4mm}, \hs{4mm}
\Phi_D = \left| \begin{array}{c} 152^o \\ 160^o \\ 170^o \end{array}
\right|
\label{eq:3-12}
\eeq

\large
2). \normalsize
In the second case $f_{0} \ra \pi^0 \pi^0$, we call $E_1$ and $E_2$
 the energies of the $\pi^0$'s and $E_3$  the energy of the $\pi^{+}$.
The general form of $F_{RES}(E_1, E_2)$ is
\beq
F_{RES}(E_1, E_2) \hs{2mm} = \hs{2mm}
D_N \hs{2mm} BW(E_3) \hs{3mm}, \label{eq:3-13}
\eeq
and because of Eq.(\ref{eq:3-8}), $D_N = D_c/\sqrt{2}$.

\vs{3mm}
\hs{3mm} \large{ } {\bf IV. \hs{3mm} Results of the model
for $D_s^{+} \ra \pi^{+} \pi^{+} \pi^{-} $ }
\normalsize
\vs{3mm}

The results of the model for the one pion energy distributions
and for the rates are now given in the three following cases :

\hs{15mm} 1) \hs{2mm}
non resonant $\pi^{+} \pi^{+} \pi^{-}$ state : $F_{NR}$

\hs{15mm} 2) \hs{2mm}
quasi two body $f_{0} \pi^{+} \ra \pi^{+} \pi^{-} \pi^{+}$ state : $F_{RES}$

\hs{15mm} 3) \hs{2mm}
superposition of the non resonant and resonant amplitudes : $F_{NR} + F_{RES}$

\large
1). \normalsize
In the non resonant case, we have an uniform Dalitz plot and for a constant
$F_{NR}$, we get from Eqs.(\ref{eq:2-13}) and (\ref{eq:2-14}) :
\beq
G_{NR}(E) = H_{NR}(E) = F_{NR}^2 \hs{1mm}
\frac{1}{m_{D_s}} \hs{1mm}
\left\{ \frac{(E^2 - m_{\pi}^2)(m_{D_s}^2 - 3 m_{\pi}^2 - 2 m_{D_s}E)}
{m_{D_s}^2 + m_{\pi}^2 - 2 m_{D_s}E } \right\}^{1/2} \hs{3mm}.
\label{eq:4-1}
\eeq
Here $E$ stands for $E_1$, $E_2$, $E_3$ indifferently.

The energy distribution is represented on Figure 1 for $F_{NR} = 1$.
The integral $I_{NR}$ given by
\beq
I_{NR} = \frac{1}{m_{D_s}}
\int_{m_{\pi}}^{ \frac{m_{D_s}^2 - \hs{1mm} 3 m_{\pi}^2}{2 \hs{1mm} m_{D_s}}}
\hs{2mm} G_{NR}(E) \hs{2mm} dE \hs{3mm},  \label{eq:4-2}
\eeq
has the numerical value
\footnote{In the limit $m_{\pi} = 0$,
the calculation of $I_{NR}$ is trivial and the result is
$I_{NR} = 0.125 \hs{1mm} F_{NR}^2$.
Correction due to $m_{\pi}  \neq 0$ is large and of the order
of $16 \%$ in spite of the very small value of $m_{\pi}/m_{D_s} \simeq 0.07$ .}
\beq
I_{NR} = 0.1053 \hs{2mm} F_{NR}^2 \hs{3mm}. \label{eq:4-3}
\eeq
Using, for $I_{NR}$, the experimental value quoted in Table 1
in the non resonant case, we obtain
\beq
| F_{NR} | \hs{2mm} = \hs{2mm}
2.15 \hs{1mm}^{+ \hs{2mm} 0.34}_{- \hs{2mm} 0.41} \hs{3mm}. \label{eq:4-4}
\eeq

\large
2). \normalsize
Now we consider the quasi two body case $D_s^{+} \ra f_{0} + \pi^{+} \ra
\pi^{+} \pi^{-} \pi^{+}$ for which the amplitude $F_{RES}$ is given by
Eq.(\ref{eq:3-2}).
It is convenient to write
\beq
|F_{RES}(E_1, E_2)|^2 \hs{2mm} =
\hs{2mm} |D_c|^2 \hs{2mm} K_c(E_1, E_2) \hs{3mm}, \label{eq:4-5}
\eeq
where the function $K_c(E_1, E_2)$ is the sum of three terms ;
\beq
K_c(E_1, E_2) = |BW(E_1)|^2 + |BW(E_2)|^2
+ 2 Re \left\{BW(E_1)BW(E_2)^{*} \right\} \hs{3mm}. \label{eq:4-6}
\eeq
After integration of $K_c(E_1, E_2)$ over $E_2$ at fixed $E_1$, as indicated
in Eq.(\ref{eq:2-13}), we obtain the $\pi^{+}$ meson energy distribution
in the form :
\beq
G_{RES}(E_1) \hs{2mm} = \hs{2mm}
|D_c|^2 \hs{2mm} K_c^{(+)}(E_1) \hs{3mm}. \label{eq:4-7}
\eeq
The function $K_c^{(+)}(E_1)$ is  represented on Figure 2.
Clearly we see the narrow peak due to the Breit-Wigner term in $E_1$
and a quasi-constant background corresponding to the Breit-Wigner term
in $E_2$.
The interference  between the two Breit-Wigner terms gives a very
small contribution to $K_c^{(+)}(E_1)$.

Integrating now $K_c(E_1, E_2)$ over $E_2$ at fixed $E_3$,
as indicated in Eq.(\ref{eq:2-14}), we obtain the $\pi^{-}$ meson
energy distribution  :
\beq
H_{RES}(E_3) \hs{2mm} = \hs{2mm}
|D_c|^2 \hs{2mm} K_c^{(-)}(E_3) \hs{3mm}. \label{eq:4-8}
\eeq
The function $K^{(-)}_c(E_3)$ is also represented on Figure 2.
The result is a quasi-constant plateau in most of
the allowed phase space domain of $E_3$.

In order to obtain the rate, we must perform a second energy integration.
The quantity $I_{RES}$ is written in the form :
\beq
I_{RES} \hs{2mm} = \hs{2mm}
|D_c|^2 \hs{2mm} K_c \hs{3mm},    \label{eq:4-9}
\eeq
where $K_c$ is given by :
\beq
K_c = \frac{1}{m_{D_s}}
\int_{m_{\pi}}^{ \frac{m_{D_s}^2 - \hs{1mm} 3 m_{\pi}^2 }
{2 \hs{1mm} m_{D_s}} } \
\hs{2mm} K_c^{(+)}(E_1) \hs{2mm} dE_1
\hs{2mm} = \hs{2mm}
\frac{1}{m_{D_s}}
\int_{m_{\pi}}^{ \frac{ m_{D_s}^2 - \hs{1mm} 3 m_{\pi}^2 }
{2 \hs{1mm} m_{D_s} } } \
\hs{2mm} K_c^{(-)}(E_3) \hs{2mm} dE_3  \hs{3mm}.
\label{eq:4-10}
\eeq
\vs{2mm}
The numerical value obtained for $K_c$ is $K_c = 5.7217$.

\vs{2mm}
Using now the results of Eq.(\ref{eq:3-12}) for $|D_c|$,
we obtain the numerical value in our model of $I_{RES}$ defined
in Eq.(\ref{eq:4-9}).
The result is
\beq
I_{RES} \hs{2mm} = \hs{2mm}
0.411 \hs{1mm}^{+ \hs{2mm} 0.128}_{- \hs{2mm} 0.186} \hs{3mm}, \label{eq:4-11}
\eeq
where, as previously explained, the errors correspond to
$\G_{\pi^{'}} = 400 \pm 200 MeV$.

The theoretical value of $I_{RES}$ in Eq.(\ref{eq:4-11}) is consistent
with the measured value of $0.481 \pm 0.192$ quoted in Table 1.
If we consider seriously the experimental one standard deviation limit,
$I_{RES} \geq 0.289$, we get a lower bound for $\G_{\pi^{'}}$,
$\G_{\pi^{'}} \geq 263 MeV$ (assuming $f_{\pi^{'}} = 40 MeV$).
There is no upper bound constraint on $\G_{\pi^{'}}$.

\large
3). \normalsize
Finally we consider the full amplitude written in Eq.(\ref{eq:3-1}).
As the constant $F_{NR}$ is real, we get
\beq
|F(E_1, E_2)|^2 = F_{NR}^2 + 2 \hs{1mm} F_{NR}
\hs{1mm} Re \left\{F_{RES}(E_1, E_2) \right\} + |F_{RES}(E_1, E_2)|^2
\hs{3mm}. \label{eq:4-12}
\eeq
We integrate over $E_1$ and $E_2$ in order to obtain the quantity $I$
of Eq.(\ref{eq:2-10}).

Using the previous results Eqs.(\ref{eq:4-3}) and (\ref{eq:4-9}), we obtain
\beq
I = 0.1053 \hs{2mm} F_{NR}^2
+ 2 \hs{2mm} F_{NR} \hs{2mm} |D_c| \hs{2mm} |J| \hs{2mm} Cos(\Phi_D + \Phi_J)
+ 5.7217 \hs{2mm} |D_c|^2 \hs{3mm}, \label{eq:4-13}
\eeq
where the complex integral $J$ is defined by :
\beq
J = \frac{1}{m_{D_s}^2}  \int \int
\hs{2mm} \left\{ BW(E_1) + BW(E_2) \right\} \hs{2mm} dE_1 \hs{2mm} dE_2
\hs{2mm} = \hs{2mm} |J| \hs{2mm} e^{i \Phi_J} \hs{3mm}.
\label{eq:4-14}
\eeq
The results are
\beq
|J| = 0.2677 \hs{10mm}, \hs{10mm} \Phi_{J} = 89.60^{0} \hs{3mm}.
\label{eq:4-15}
\eeq

Using the experimental constraints for $I_{NR}$ and $I$ given in Table 1,
from Eqs.(\ref{eq:4-3}) and (\ref{eq:4-13}), we can check the consistency
of our model by determining the parameter $F_{NR}$
and compare to Eq.(\ref{eq:4-4}).
The results are presented on Table 2 for three value of $\G_{\pi^{'}}$,
$\G_{\pi^{'}} = 600, 400, 263 \hs{2mm} MeV$.
For large value of $\G_{\pi^{'}}$,
we obtain only positive solutions for $F_{NR}$.
When $\G_{\pi^{'}}$ decreases,  $|D_c|$ decreases and
the interference term in Eq.(\ref{eq:4-13}) becomes smaller,
then we obtain both positive and negative solutions for $F_{NR}$.

\vs{3mm}
\begin{table}[thb]
\begin{center}
\begin{tabular}{|c||c||c||c||c|}
\hline
$\G_{\pi^{'}}$ (MeV) &
$F_{NR}$ &
$I_{NR}$ &
$I$  &
$I_{RES}$ \\
\hline
\hline
600 &
1.7378 \hs{10mm} 1.9898 &
0.318 \hs{10mm} 0.417 &
0.719 \hs{10mm} 0.798 &
0.5385 \\
\hline
400 &
1.7378 \hs{10mm} 2.1680 &
0.318 \hs{10mm} 0.495 &
0.642 \hs{10mm} 0.798 &
0.4105 \\
\hline
263 &
1.7378 \hs{10mm} 2.3385 &
0.318 \hs{10mm} 0.576 &
0.557 \hs{10mm} 0.798 &
0.2890 \\
& -1.7378 \hs{8mm} -2.0671 &
0.318 \hs{10mm} 0.456 &
0.657 \hs{10mm} 0.798 &
0.2890 \\
\hline
Experiment &
$-$ &
0.318 $ - $ 0.654 &
0.500 $ - $ 0.798 &
0.289 $ - $ 0.673 \\
\hline
\hline
\end{tabular}

\vs{3mm}
Table 2 \\
\end{center}
\end{table}

The full pion energy distributions
coming from the total amplitude in Eq.(\ref{eq:4-12})
depend on the two quantities
$F_{NR}$ and $D_c$.
Writing the squared  modulus of the total amplitude in the form :
\beq
|F(E_1, E_2)|^2 = F_{NR}^2
+ 2 F_{NR} \hs{2mm} Re \left\{ D_c \hs{1mm} J_c(E_1, E_2) \right\}
+ |D_c|^2 \hs{1mm} K_c(E_1, E_2)  \hs{3mm}, \label{eq:4-16}
\eeq
where $K_c(E_1, E_2)$ is given by Eq.(\ref{eq:4-6})
and $J_c(E_1, E_2)$ is simply given by :
\beq
J_c(E_1, E_2) = BW(E_1) + BW(E_2) \hs{3mm}. \label{eq:4-17}
\eeq
Of course, $K_c(E_1, E_2) = |J_c(E_1, E_2)|^2$.

After integration of Eq.(\ref{eq:4-16})
over $E_2$ at fixed $E_1$, we obtain the $\pi^{+}$
meson energy distribution $G_c(E_1)$ :
\beq
G_c(E_1) = F_{NR}^2 \hs{1mm} G_{NR}(E_1) + 2 \hs{2mm} F_{NR} \hs{2mm}
Re \left\{D_c \hs{1mm} J_c^{+}(E_1) \right\}
+ |D_c|^2 \hs{1mm} K_c^{+}(E_1) \hs{3mm}.  \label{eq:4-18}
\eeq
The quantities $G_{NR}(E_1)$ and $K_c^{+}(E_1)$ have been represented
on Figures 1 and 2, respectively, and $J_c^{+}(E_1)$ is defined by :
\beq
J_c^{+}(E_1) = \frac{1}{m_{D_s}} \hs{2mm} \int_{E_{-}(E_1)}^{E_{+}(E_1)}
\hs{2mm} J_c(E_1, E_2) \hs{2mm} dE_2 \hs{3mm}.  \label{eq:4-19}
\eeq

In a similar way, integrating over $E_2$ at fixed $E_3$, we obtain
the $\pi^{-}$ meson energy distribution in the form :
\beq
H_c(E_3) = F_{NR}^2 \hs{1mm} G_{NR}(E_3) + 2 \hs{2mm} F_{NR} \hs{2mm}
Re \left\{D_c \hs{1mm} J_c^{-}(E_3) \right\}
+ |D_c|^2 \hs{1mm} K_c^{-}(E_3)  \hs{3mm}. \label{eq:4-20}
\eeq
In Eq.(\ref{eq:4-20}), $G_{NR}(E_3)$ as given by Eq.(\ref{eq:4-1})
is represented on Figure 1.
The quantity $K_c^{-}(E_3)$ is represented on Figure 2
and $J_c^{-}(E_3)$ is defined by :
\beq
J_c^{-}(E_3) = \frac{1}{m_{D_s}}  \int_{E_{-}(E_3)}^{E_{+}(E_3)}
\hs{2mm} J_c(m_{D_s}-E_2-E_3, E_2) \hs{2mm} dE_2 \hs{3mm}.  \label{eq:4-21}
\eeq
As mentioned above, the full $\pi^{+}$ and $\pi^{-}$ energy distributions
$G_c(E_1)$ and $H_c(E_3)$ are functions of $F_{NR}$ and $D_c$.
They are represented respectively in Figures 3 and 4.

\vs{3mm}
\hs{3mm} \large{ }
{\bf V. \hs{3mm} Prediction of the model for $D_s^{+} \ra \pi^{0} \pi^{0}
\pi^{+} $ }
\normalsize
\vs{3mm}

We have no experimental data on the decay mode
$D_s^{+} \ra \pi^0 \pi^0 \pi^{+}$.
However our model can make predictions for the three following cases :

\hs{15mm} 1) non resonant $\pi^0 \pi^0 \pi^{+}$ state,

\hs{15mm} 2) quasi two body $f_{0} \pi^{+} \ra  \pi^0 \pi^0 \pi^{+}$ state,

\hs{15mm} 3) superposition of the non resonant and resonant amplitude.

\large
1). \normalsize
In the non resonant case of a constant function $F_{NR}(E_1, E_2)$,
we obtain the same shape for one pion energy distribution as shown
on Figure 1.

However the constant $F_{NR}$ has no reason to be the same for the two modes
$\pi^{+} \pi^{+} \pi^{-}$ and $\pi^0 \pi^0 \pi^{-}$.
We now discuss this problem.
By assumption, with a constant function $F_{NR}(E_1, E_2)$,
we have a full symmetry in space between the three pions.
{}From the Bose-Einstein symmetry,
the isospin configuration has also to be totally symmetric.
Consider now a third rank fully symmetric tensor in a 3 dimensional space.
It has 10 independent components.
With respect to the isospin SO(3) orthogonal group,
such a tensor is reducible
into an isospin $I = 3$ part with 7 components
and an isospin $I = 1$ part with 3 components.
In our specific problem, only the later part contributes,
the $u\ol{d}$ weak current being an isovector.
By inspection of the relevant Clebsch-Gordan coefficients, we obtain
the result :
\beq
F_{NR}(\pi^{+}\pi^{+}\pi^{-})
= 3 \hs{2mm} F_{NR}(\pi^{0}\pi^{0}\pi^{+}) \hs{3mm}. \label{eq:5-1}
\eeq
As a consequence, in our model, we obtain
for the non resonant part :
\beq
\frac{ \G(D_s^{+} \ra \pi^0 \pi^0 \pi^{+})_{NR}}
{ \G(D_s^{+} \ra \pi^{+} \pi^{+} \pi^{-})_{NR}}
= 11.1 \hs{2mm} \% \hs{3mm}, \label{eq:5-2}
\eeq
and the non resonant branching ratio
$Br(D_s^{+} \ra \pi^0 \pi^0 \pi^{+})_{NR}$ is
expected to occur only at the $10^{-3}$ level.

\large
2). \normalsize
For the quasi two body decay of
$D_s^{+} \ra f_{0} \pi^{+} \ra \pi^0 \pi^0 \pi^{+}$,
the amplitude $F_{RES}$ is given by Eq.(\ref{eq:3-13}) :
\beq
|F_{RES}(E_1, E_2)|^2 \hs{2mm} = \hs{2mm}
|D_N|^2 \hs{2mm} K_N(E_1, E_2) \label{eq:5-3}
\eeq
where the function $K_N(E_1, E_2)$ is simply given by :
\beq
K_N(E_1, E_2) \hs{1mm} = \hs{1mm} |BW(E_3)|^2
\hs{1mm} = \hs{1mm} |BW(m_{D_s}-E_1-E_2)|^2  \label{eq:5-4}
\eeq
After integration over $E_2$ at fixed $E_1$,
as indicated in Eq.(\ref{eq:2-13}),
we obtain the $\pi^0$ meson energy distribution in the form :
\beq
G_{RES}(E_1) \hs{2mm} = \hs{2mm}
|D_N|^2 \hs{2mm} K_N^{(0)}(E_1)   \label{eq:5-5}
\eeq
The function $ K_N^{(0)}(E_1)$ is a quasi flat distribution
represented on Figure 5.

Integrating now over $E_2$ at fixed $E_3$,
as indicated in Eq.(\ref{eq:2-14}),
we have the $\pi^{+}$ energy distribution in the form :
\beq
H_{RES}(E_3) \hs{2mm} = \hs{2mm}
|D_N|^2 \hs{2mm} K_N^{(+)}(E_3)   \label{eq:5-6}
\eeq
The function $K_N^{(+)}(E_3)$ is represented on Figure 5 and
the result exhibits a narrow Breit-Wigner peak similar to
the one drawn on Figure 2 for $K^{+}_c(E_1)$.

The decay rate is now obtained by performing a second energy integration.
The quantity $I_{RES}$ is written in the form :
\beq
I_{RES} \hs{2mm} = \hs{2mm} |D_N|^2 \hs{2mm} K_N \hs{3mm},  \label{eq:5-7}
\eeq
where $K_N$ is given by :
\beq
K_N = \frac{1}{m_{D_s}} \ \hs{1mm}
\int_{m_{\pi}}^{ \frac{m_{D_s}^2 - \hs{1mm} 3 m_{\pi}^2 }
 {2 \hs{1mm} m_{D_s}} } \
\hs{2mm} K_N^{(0)}(E_1) \hs{2mm} dE_1
\hs{2mm} = \hs{2mm}
\frac{1}{m_{D_s}} \hs{1mm}
\int_{m_{\pi}}^{ \frac{ m_{D_s}^2 - \hs{1mm} 3 m_{\pi}^2 }
{2 \hs{1mm} m_{D_s} } } \
\hs{2mm} K_N^{(+)}(E_3) \hs{2mm} dE_3
\label{eq:5-8}
\eeq
The numerical value of $K_N$ is $K_N = 2.7469$.

Using the relation between $D_N$ and $D_c$, we make the prediction
\beq
\frac{ \G(D_s^{+} \ra f_{0} \pi^{+} \ra \pi^0 \pi^0 \pi^{+})}
{ \G(D_s^{+} \ra f_{0} \pi^{+} \ra \pi^{+} \pi^{-} \pi^{+})}
\hs{2mm} = \hs{2mm} 24 \hs{1mm} \%   \label{eq:5-9}
\eeq
The departure of this ratio from $ 25 \hs{2mm} \% $ is simply
due to the interference
between the two Breit-Wigner contributions in
$D_s^{+} \ra f_{0} \pi^{+} \ra \pi^{+} \pi^{-} \pi^{+}$.

\large
3). \normalsize
We finally consider the full amplitude written in the form
of Eq.(\ref{eq:3-1}).
An equation similar to Eq.(\ref{eq:4-13}) gives the total rate,
and going from the mode $\pi^{+} \pi^{+} \pi^{-}$
to the mode $\pi^0 \pi^0 \pi^{+}$ implies the changes :
\beq
F_{NR} \ra \frac{1}{3} F_{NR} \hs{5mm}, \hs{5mm}
D_c \ra \frac{1}{\sqrt{2}} D_c \hs{5mm}, \hs{5mm}
J \ra \frac{1}{2} J \hs{5mm}, \hs{5mm}
K_c \ra K_N    \label{eq:5-10}
\eeq
and we obtain instead of Eq:(\ref{eq:4-13}) :
\beq
I_{\pi^0 \pi^0 \pi^{+}} \hs{2mm} = \hs{2mm}
0.0117 F_{NR}^2
+ 0.2357 F_{NR} |D_c| |J| Cos(\Phi_D + \Phi_J)
+ 1.3735 |D_c|^2 \hs{2mm}.  \label{eq:5-11}
\eeq

Using the solutions for $F_{NR}$ given in the Table 2, we can compute
the decay rate for the $\pi^0 \pi^0 \pi^{+}$ mode
as a function of $F_{NR}$ and of the $\pi^{'}$ width $\G_{\pi^{'}}$
via the quantity $D_c$.
In Table 3, we show the results for the ratio $R$ defined by :
\beq
R \hs{2mm} = \hs{2mm}
\frac{ \G(\pi^0 \pi^0 \pi^{+})}
{ \G(\pi^{+} \pi^{+} \pi^{-})} \hs{2mm}.  \label{eq:5-12}
\eeq

\vs{3mm}
\begin{table}[thb]
\begin{center}
\begin{tabular}{|c||c||c|}
\hline
$\G_{\pi^{'}}$ (MeV) &
$F_{NR}$ &
$R$  \\
\hline
\hline
600 &
1.7378 \hs{10mm} 1.9898 &
$20.2 \pm 0.5 \hs{2mm} \%$ \\
\hline
400 &
1.7378 \hs{10mm} 2.1680 &
$18.5 \pm 0.8 \hs{2mm} \%$ \\
\hline
263 &
1.7378 \hs{10mm} 2.3385 &
$16.7 \pm 1 \hs{2mm} \% $ \\
& -1.7378 \hs{8mm} -2.0671 &
$16.3 \pm 0.5 \hs{2mm} \% $ \\
\hline
\hline
\end{tabular}

\vs{3mm}
Table 3 \\
\end{center}
\end{table}

The pion energy distributions depend on $F_{NR}$ and $D_c$,
and we have for the squared amplitude an expression similar to
Eq.(\ref{eq:4-16}) with the appropriated changes
 indicated on Eq.(\ref{eq:5-10}) :
\beq
|F(E_1, E_2)|^2 \hs{2mm} = \hs{2mm}
\frac{1}{9} F_{NR}^2
+ \frac{2}{3 \sqrt{2}} F_{NR} Re \left\{D_c J_N(E_1, E_2) \right\}
+ \frac{1}{2} |D_c|^2 K_N(E_1, E_2) \hs{2mm}, \label{eq:5-13}
\eeq
where $J_N(E_1, E_2)$ contains only one Breit-Wigner amplitude :
\beq
J_N(E_1, E_2) \hs{2mm} = \hs{2mm}
BW(E_3) \hs{2mm} = \hs{2mm}
BW(m_{D_s}-E_1-E_2) \hs{2mm}, \label{eq:5-14}
\eeq
and $K_N(E_1, E_2) = |J_N(E_1, E_2)|^2$.

The full $\pi^{0}$ energy distribution is obtained after integration of
$|F(E_1, E_2)|^2$ over $E_2$ at fixed $E_1$,
and the full $\pi^{+}$ energy distribution
after intergration over $E_2$ at fixed $E_3$ :
\beqa
G_N(E_1) \hs{2mm} &=& \hs{2mm}
\frac{1}{9} F_{NR}^2 G_{NR}(E_1)
+ \frac{2}{3 \sqrt{2}} F_{NR} Re \left\{D_c J_N^{(0)}(E_1) \right\}
+ \frac{1}{2} |D_c|^2 K_N^{(0)}(E_1) \hs{2mm}, \label{eq:5-15} \\
H_N(E_3) \hs{2mm} &=& \hs{2mm}
\frac{1}{9} F_{NR}^2 G_{NR}(E_3)
+ \frac{2}{3 \sqrt{2}} F_{NR} Re \left\{D_c J_N^{(+)}(E_3) \right\}
+ \frac{1}{2} |D_c|^2 K_N^{(+)}(E_3) \hs{2mm}, \label{eq:5-16}
\eeqa
The functions $K_N^{(0)}(E_1)$ and $K_N^{+}(E_3)$
have been represented on Figure 5.
The functions $J_N^{(0)}$ and $J_N^{(+)}$ are defined by :
\beqa
J_N^{(0)}(E_1) \hs{2mm} &=& \hs{2mm}
\frac{1}{m_{D_s}} \int_{E_{-}(E_1)}^{E_{+}(E_1)}
BW(m_{D_s}-E_1-E_2) \hs{1mm} dE_2 \hs{2mm}, \label{eq:5-17} \\
J_N^{(+)}(E_3) \hs{2mm} &=& \hs{2mm}
\frac{1}{m_{D_s}} \int_{E_{-}(E_3)}^{E_{+}(E_3)}
BW(E_3) \hs{1mm} dE_2 \hs{2mm}, \label{eq:5-18}
\eeqa
The $\pi^{0}$ and $\pi^{+}$ energy distributions
$G_N(E_1)$ and $H_N(E_3)$ are functions of $F_{NR}$ and $D_c$.
They are represented on Figures 6 and 7 with the same values
for $\G_{\pi^{'}}$ and $F_{NR}$ as those used in Figures 3 and 4.
The first observation is the minor role now played
by the constant non resonant component essentially
due to the isospin factor Eq.(\ref{eq:5-1}).
However the sign of the interference between the two components
can be observed at large values of $E_3$
as illustrated in Figures 7-a and 7-b.

\vs{5mm}
\hspace{3mm} \large{} {\bf V. \hs{3mm} Summary and Concluding Remarks}
\vspace{3mm}
\normalsize

The large value experimentally observed for the branching ratio
of the decay mode $D_s^{+} \ra \pi^{+} \pi^{+} \pi^{-}$
is a very interesting problem which necessitates a better understanding
of the role played by the $W$ annihilation mechanism
in $D$ meson  decay, and which is probably more important
than usually expected.

We have proposed to construct the unique structure function
$F(E_1, E_2)$ as the sum of a constant non resonant term
described by a real parameter $F_{NR}$, and a Breit-Wigner resonant term
associated to a quasi two body $f_{0}(980) + \pi^{+}$ state
which seems to play an important role
as indicated by experiments.
The quasi two body $\rho^{0} + \pi$ state which is commonly
considered by previous authors \cite{R2, R3, R6}
is experimentally depressed and has been disregarded here.

It is possible to determine the resonant amplitude using parameters
experimentally known, even if a large uncertainty remains on their
precise experimental values.
We have only retained, for simplicity, the uncertainty
due to the $\pi(1300)$ width.
Our model depends only on one free parameter,
and we show that it is possible to find values of $F_{NR}$ fitting
the three observed rates : total $D_s^{+} \ra \pi^{+} \pi^{+} \pi^{-}$,
non resonant $(D_s^{+} \ra \pi^{+} \pi^{+} \pi^{-})_{NR}$, and
resonant $D_s^{+} \ra f_{0} \pi^{+} \ra \pi^{+} \pi^{+} \pi^{-}$.
This result is obviously non trivial.

Of course, the Dalitz plots are determined by
the squared modulus of the structure function $|F(E_1, E_2)|^2$
and they are predicted by our model.
Before the double differential quantities could be observed
for the reason of statistics,
it would be easier to measure the one pion
energy distributions which are nothing but projections of the Dalitz plots
on one energy axis.
The figures shown in this paper present the various situations
to be compared with experimental results when available.

The decay mode $D_s^{+} \ra \pi^0 \pi^0 \pi^{+}$ has not been experimentally
observed,
however the rates, the pion energy distributions, and the Dalitz plots
can be computed in our model.
We compare the $\pi^0 \pi^0 \pi^{+}$ and
$\pi^{+} \pi^{+} \pi^{-}$ integrated widths
through the ratio $R$ in Eq.(\ref{eq:5-12}) and Table 3,
and we also draw the $\pi^0$ and $\pi^{+}$ meson energy distributions.
The quantity $R$ is predicted to be in the $16 \% - 20 \%$ range
and that might be the reason why the mode $D_s^{+} \ra \pi^0 \pi^0 \pi^{+}$
has not yet been detected,
its branching ratio being expected to be few $10^{-3}$.

We wish to emphasize that in the absence of additional
 experimental informations,
our crude model is the simplest one accounting for the observed rates.
In particular, for the resonant component,
the normalization is estimated theoretically and constrained experimentally.
Fortunately both approaches overlap,
a fact a priori not evident, and they can be improved
when experimental uncertainty will be reduced.

\vspace{10mm}
\hspace{1cm} \Large{} {\bf Acknowledgements}    \vspace{0.5cm}
\normalsize

\vs{2mm}
Y. Y. K would like to thank the Commissariat \`a l'Energie Atomique of France
for the award of a fellowship
and especially G. Cohen-Tannoudji and J. Ha\"{\i}ssinski
for their encouragements.

%
%
%
%

\vs{10mm}
%

\newpage
\section*{Figure captions}
\normalsize
\vspace{0.5cm}

\ben

\item
{\bf Figure 1} : \hs{3mm}
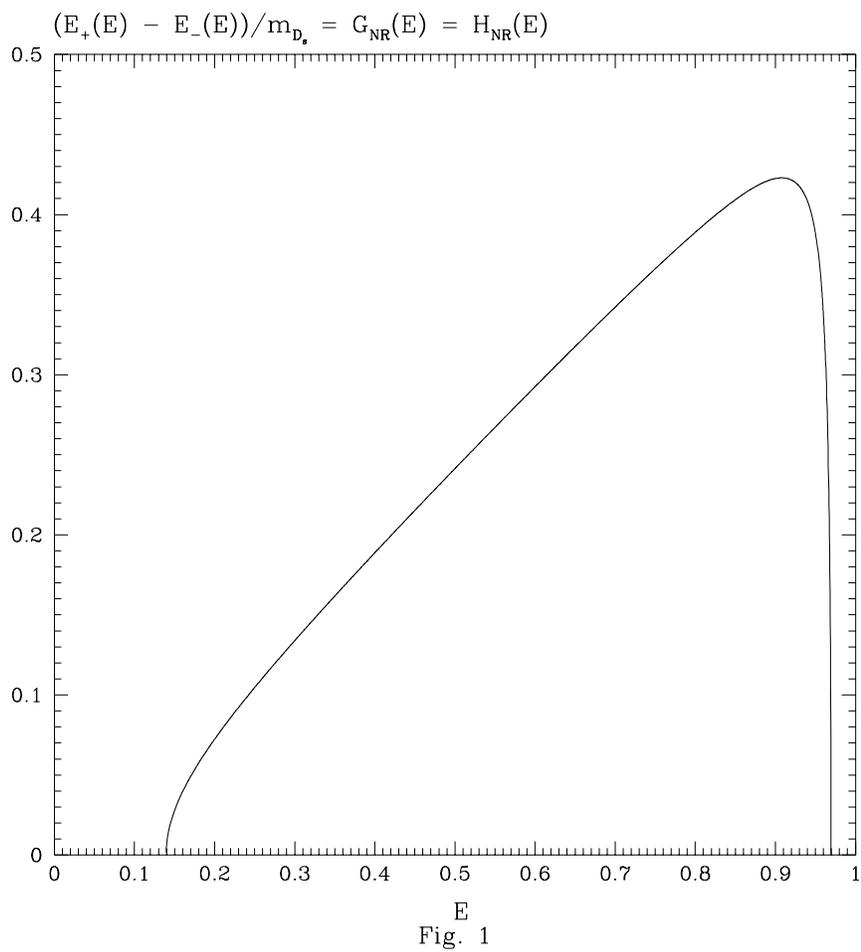
The shape of the pion energy distributions for a uniform Dalitz plot.
$E$ stands for $E_1$, $E_2$, $E_3$ indifferently.
\item
{\bf Figure 2} : \hs{3mm}
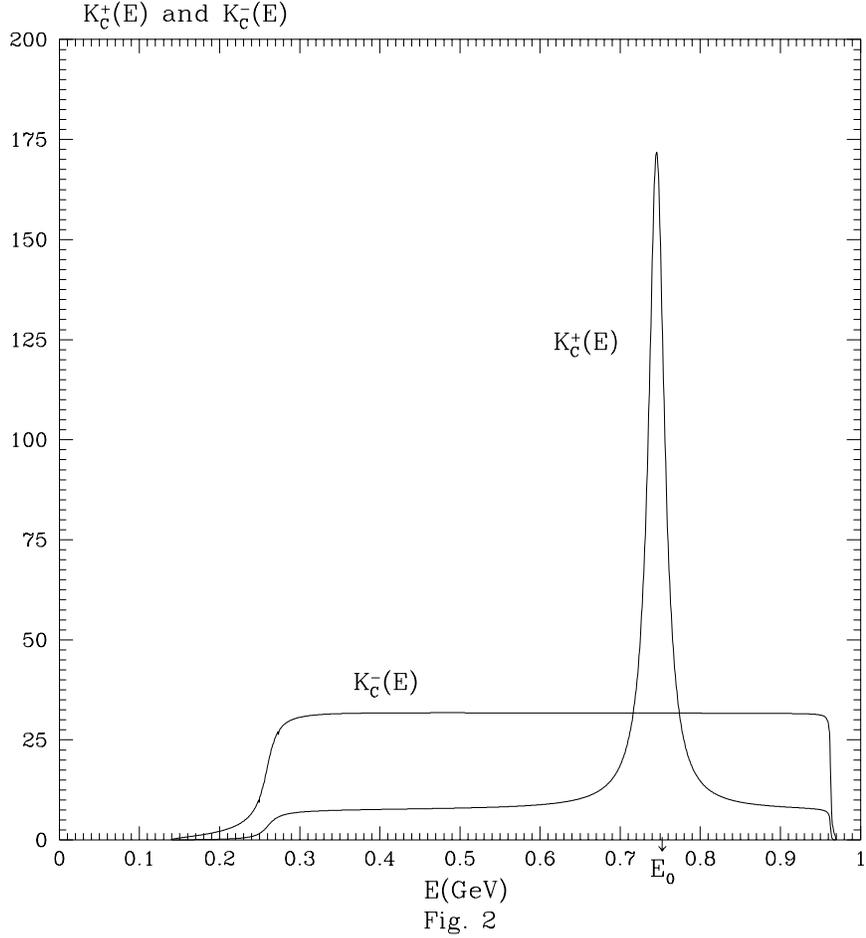
The shapes of the $\pi^{+}$ and $\pi^{-}$ meson energy distributions
for $D_s^{+} \ra f_{0} \pi^{+} \ra \pi^{+} \pi^{-} \pi^{+}$.
$K_c^{+}(E)$ represents the energy distribution of $\pi^{+}$ meson and
$K_c^{-}(E)$ represents that of $\pi^{-}$ meson.
The quantity $E_0 = 7.45 \hs{2mm} GeV$ is associated to the $f_{0}$ resonance.

\item
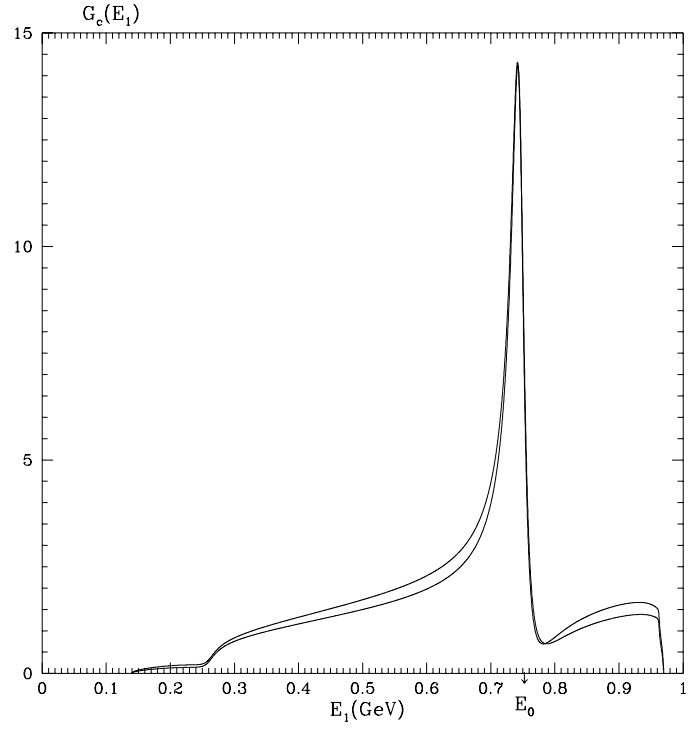
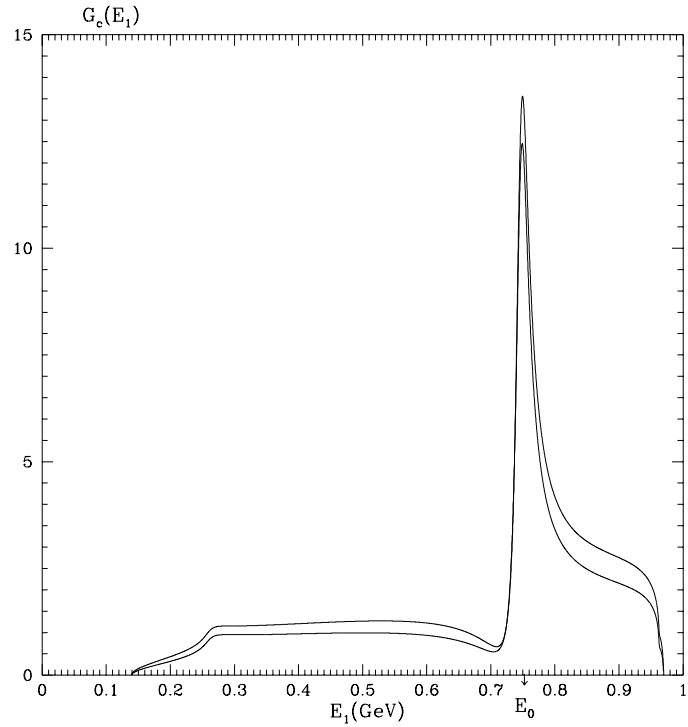
{\bf Figure 3} : \hs{3mm}
The fully normalized
$\pi^{+}$ meson energy distribution
for $D_s^{+} \ra \pi^{+} \pi^{-} \pi^{+}$.

\hs{10mm} 3-a) for $\G_{\pi^{'}} = 600 \hs{2mm} MeV$,
\hs{14mm} $1.74 \leq F_{NR} \leq 1.99$

\hs{10mm} 3-b) for $\G_{\pi^{'}} = 263 \hs{2mm} MeV$,
\hs{10mm} $-2.07 \leq F_{NR} \leq  -1.74$

The two lines in each curve correspond to the extremum values of $F_{NR}$.
The quantity $E_0 = 7.45 \hs{2mm} GeV$ is associated to the $f_{0}$ resonance.

\item
{\bf Figure 4} : \hs{3mm}
The fully normalized
$\pi^{-}$ meson energy distribution
for $D_s^{+} \ra \pi^{+} \pi^{-} \pi^{+}$. \\
The parameters are the same as in Figure 3.

\item
{\bf Figure 5} : \hs{3mm}
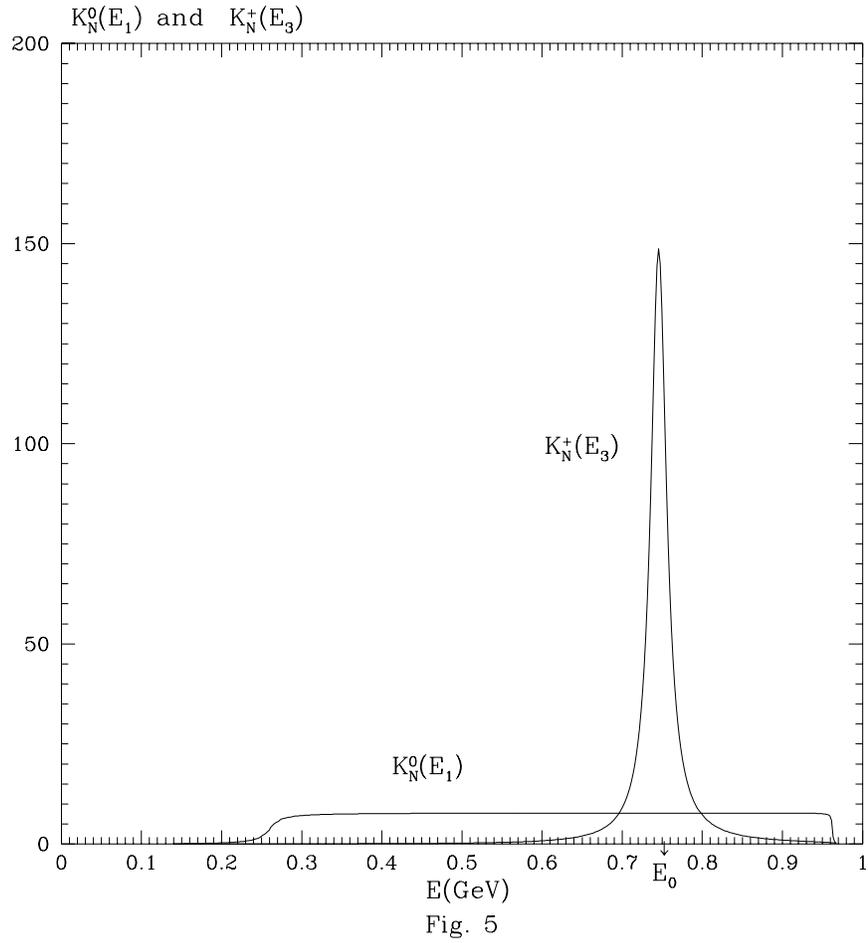
The shapes of the $\pi^{0}$ and $\pi^{+}$ meson energy distributions
for $D_s^{+} \ra f_{0} \pi^{+} \ra \pi^{0} \pi^{0} \pi^{+}$.
The quantity $E_0 = 7.45 \hs{2mm} GeV$ is associated to the $f_{0}$ resonance.

\item
{\bf Figure 6} : \hs{3mm}
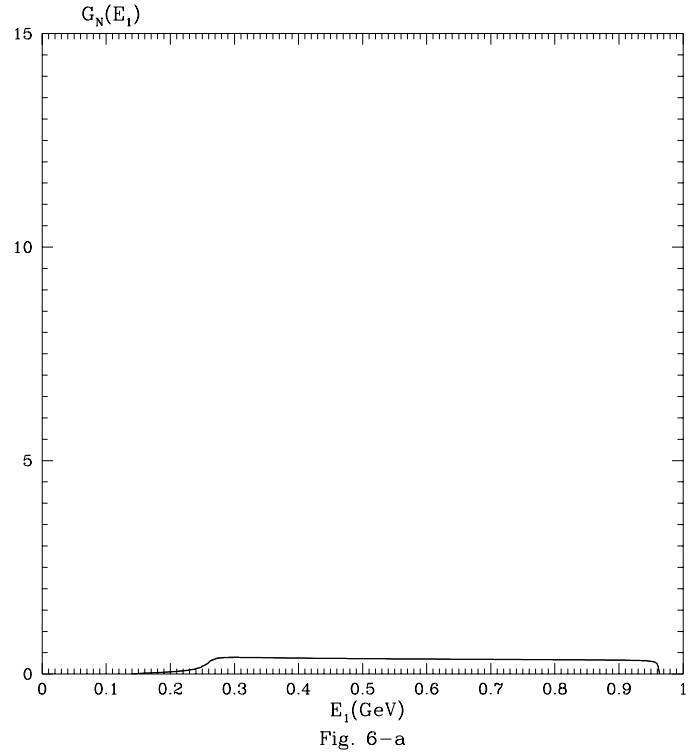
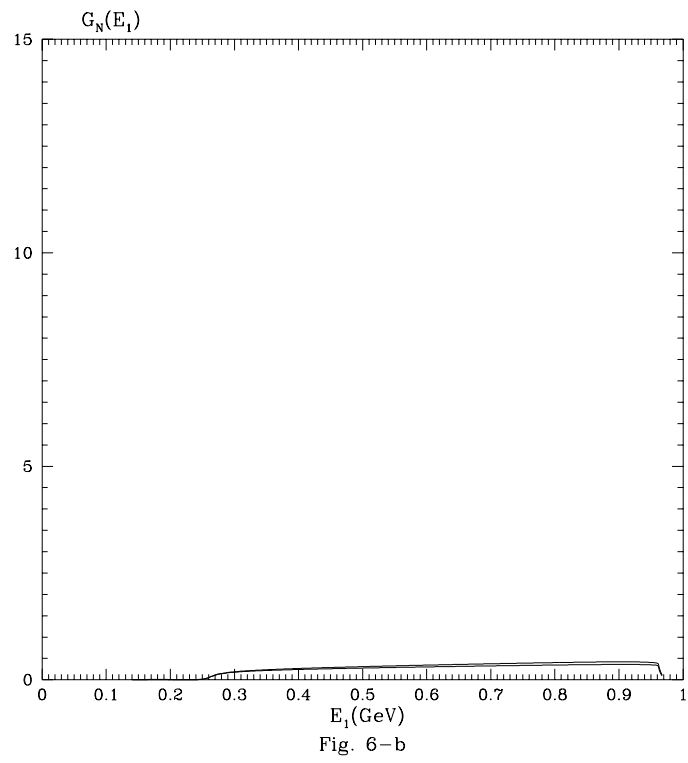
The fully normalized
$\pi^{0}$ meson energy distribution
for $D_s^{+} \ra \pi^{0} \pi^{0} \pi^{+}$. \\
The parameters are the same as in Figure 3.

\item
{\bf Figure 7} : \hs{3mm}
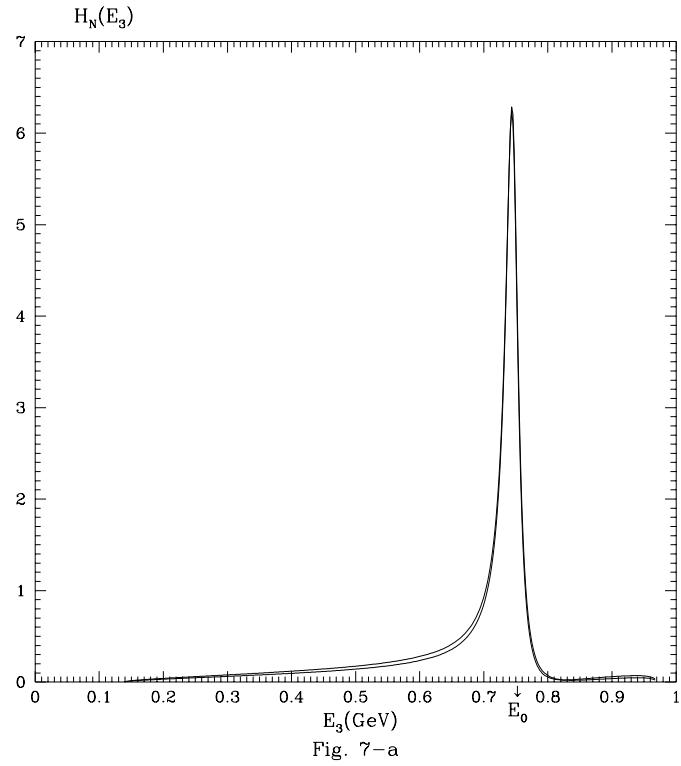
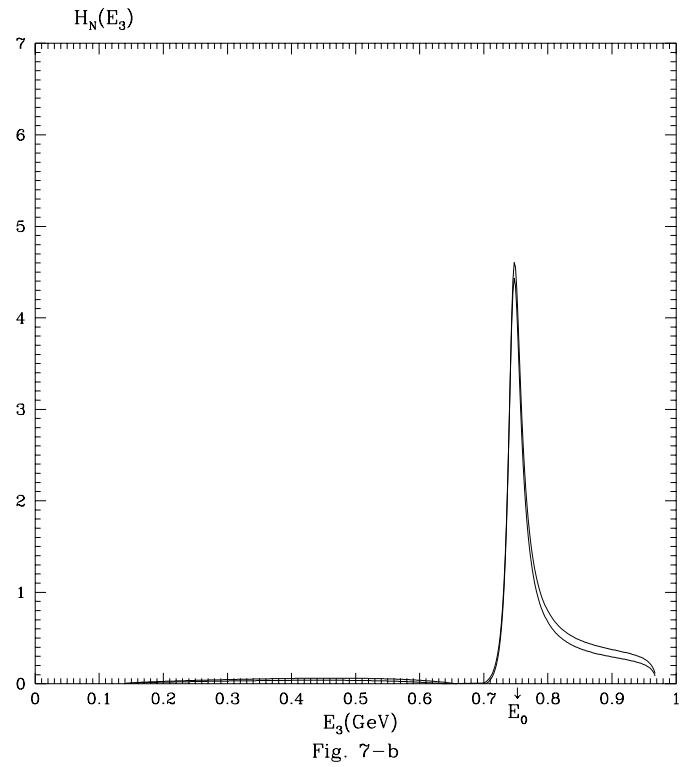
The fully normalized
$\pi^{+}$ meson energy distribution
for $D_s^{+} \ra \pi^{0} \pi^{0} \pi^{+}$. \\
The parameters are the same as in Figure 3.
The quantity $E_0 = 7.45 \hs{2mm} GeV$ is associated to the $f_{0}$ resonance.

\newpage

\een


\begin{thebibliography}{99}
%
\bibitem{R1}
Review of Particles Properties, Reference as Phys. Rev. {\bf D 50},
August 1994.
%
\bibitem{R2}
J. H. K\"uhn and E. Mirkes, {\em Phys. Lett.} {\bf B 286} 381 (1992);
{\em Z. Phys.} {\bf C 56} 661 (1992).
%
\bibitem{R3}
J. Stern, N. H. Fuchs and M. Knecht, Proceedings of the 3rd Workshop on the
$\tau$-Charm factory, Preprint IPN/TH 93-38.
%
\bibitem{R4}
M. ~Wirbel, B. ~Stech and M. ~Bauer, {\em Z. Phys.} {\bf C29} 637 (1985); \\
M. ~Bauer, B. ~Stech and M. ~Wirbel, {\em Z. Phys.} {\bf C34} 103 (1987).
%
\bibitem{R5}
M. Neubert, V. Rieckert, B. Stech and Q. P. Xu,
Advanced Series on Directions in High Energy Physics Vol. 10,
$\ul{\bf \rm Heavy \hs{2mm} Flavours}$, page
286, Editors A.J. Buras and M. Lindner,
(World Scientific, Singapore, 1992).
%
\bibitem{R6}
N. Isgur, C. Morningstar and C. Reader, {\em Phys.  Rev.} {\bf D 39} 1357
(1989)

\end{thebibliography}
\end{document}